\documentclass[aps,prl,twocolumn,superscriptaddress,oneside,floatfix,amsmath,showpacs,amssymb]{revtex4-1}

\usepackage{graphicx}
\usepackage{dcolumn}
\usepackage{bm}
\usepackage[usenames]{color}
\usepackage{hyperref}
\usepackage[normalem]{ulem}
\begin{document}

\bibliographystyle{unsrt}

\title{Anharmonic coupling between electrons and TO phonons in the vicinity of ferroelectric quantum critical point.}%Enhanced thermoelectric conductivity due to electron-phonon coupling in an anharmonic lattice}%Anharmonic electron-phonon coupling due to collective fluctuations close to ferroelectric quantum phase transition.}

\author{P.Chudzinski}
\affiliation{ASC, School of Mathematics and Physics, Queen's University Belfast}

\date{\today}

\begin{abstract}

We explore a novel coupling mechanism of electrons with the transverse optical (TO) phonon branch in a regime when the TO mode becomes highly anharmonic and drives the ferroelectric phase transition. We show that this anharmonicity, which leads to a collective motion of ions, is able to couple electronic and lattice displacement fields. An effective correlated electron-ion dynamics method is required to capture the effect of the onset of the local electric polarization due to this collective behavior close to the quantum critical point. We identify an intermediate temperature range where an emergent phonon drag may contribute substantially to thermoelectric conductivity in this regime. We find that, under optimal conditions, this extra contribution may be larger than values achieved so far in the benchmark material, PbTe. In the last part we make a case for the importance of our results in the generic problem of anharmonic electron-lattice dynamics.   %\emph{, hence our study brings a promising, yet unexplored pathway towards a long sought improvement of thermoelectric efficiency.}  

% -- its softening drives the system to the transition. This softening results in an anomalously strong dispersion of the optical mode and brings it very close to acoustic branches. 
%Our study is motivated by recent experimental findings in PbTe where hybridization with the LA mode and the so-called 'waterfall' of spectral weight were detected. 
%By accounting for a strong anharmonicity present in the system close to the ferroelectric  transition, we are able to identify a novel coupling mechanism which is driven by the presence of fluctuations of the order parameter in the critical regime.  %Subsequently, we consider the coupling of TO phonons with electrons as well as TO-LA phonon-phonon scattering. Since 
%Our study is dedicated to intermediate temperatures, where the phonon drag effect is the largest, an effect which, as we show, manifest itself in thermoelectric transport coefficient.

\end{abstract}

\pacs{
}

\maketitle

%\subsubsection{introduction}

Motivated by the search for new efficient renewable sources of energy, a huge experimental effort has been made towards identifying materials with better thermoelectric properties in the recent years. So far, this intense search has been based mostly on a single particle theoretical description of electrons and standard electron-phonon coupling due to displacement potential, supported by numerical ab-initio simulations\citep{Toborai-band-eng}\citep{Pohls-abinit}. While this research program has had some notable successes\citep{Snyder-cmplx-thermoel} it should be also noted that the progress has been slow and so novel pathways are needed. At the same time it has been observed that in several cases good thermoelectrics (TE) are weakly doped semiconductors, such as PbTe\citep{Snyder-band-eng}, SnTe or SrTiO$_3$, that are in the vicinity of a ferroelectric quantum critical phase transition (FE-QCP). It is then natural to ask if it is just a pure coincidence, or whether an yet undiscovered mechanism enhances the Seebeck coefficient. 

As the system approaches the FE-QCP, a crucial role is played by the transverse optical (TO) phonons. In the Landau macroscopic framework the TO mode spectral weight is coupled to the material's electric polarization\citep{Devonshire-orig} (the order parameter), so as the FE transition takes place, the TO softening at $q=0$ indicates the emergence of a uniform displacement and the polar order. The electrons should be susceptible to this dipole ordering and usually
% (apart from a few exotic multi-orbital materials\citep{Capone-FE+metal})
the appearance of the FE is accompanied by a disappearance of the electron pockets. Otherwise the remaining free electrons would screen the FE order. A massive softening at the $\Gamma$ point implies that the TO branch emerges as a new family of phonons with a very substantial velocity, often comparable to the one of the longitudinal acoustic (LA) branch\citep{Zhang-anharm-PbTe}, and so they certainly contribute to the transport properties of the system. 

\begin{figure}[t]
\centering
 \includegraphics[width=1.1\linewidth]{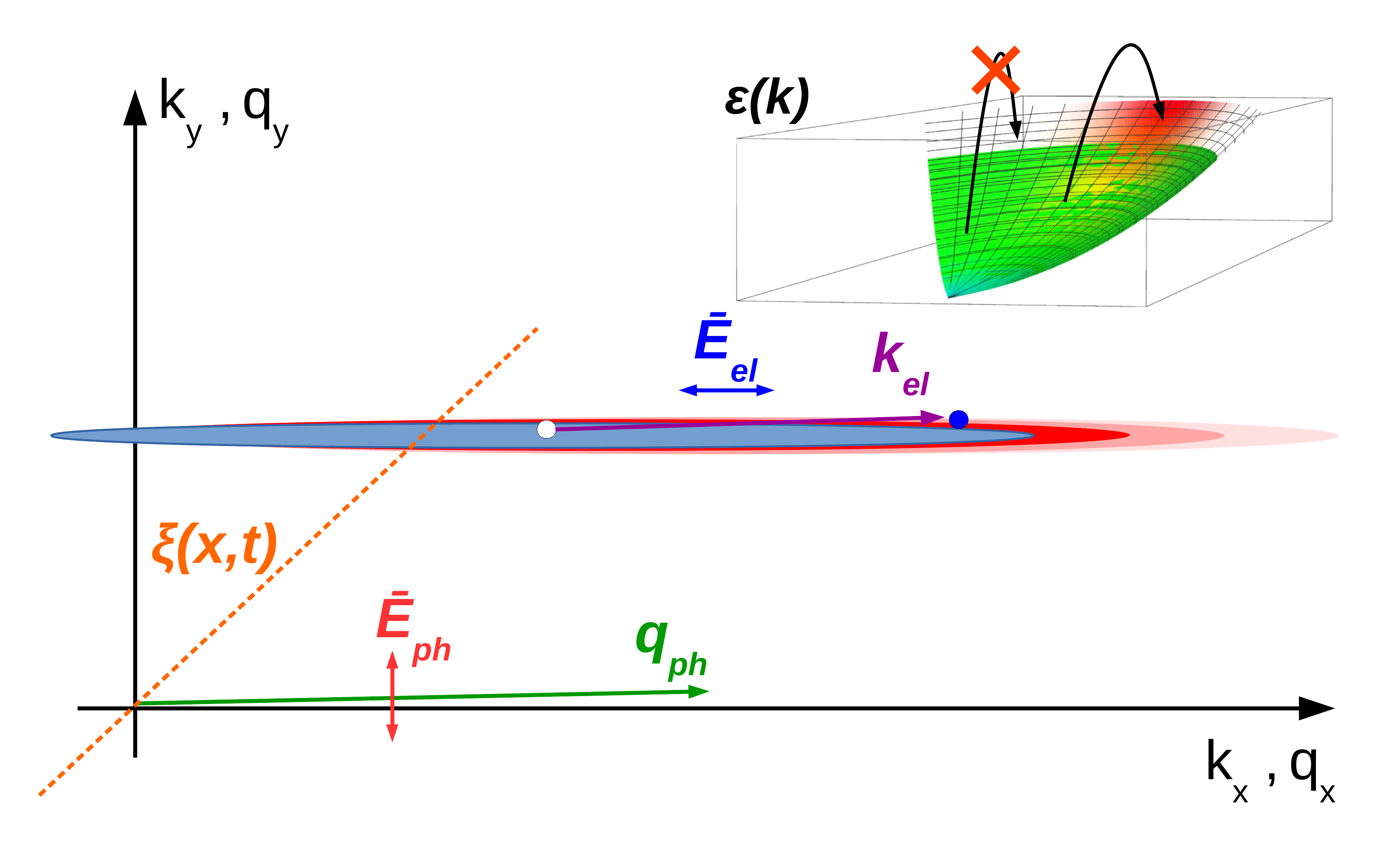}
  \caption{In the conventional model for coupling between electron and lattice displacements an electron-hole excitation in a Fermi pocket with momentum $k_{el}$ (blue) creates an electric field $\vec{E}_{el}$ while the matching TO mode with momentum $q_{ph}$ (green), $q_{ph}=k_{el}$, moves lattice atoms %(as described by its generalized variabe $\Xi^{TO} \perp q $) and 
	in a way that induces an electric field $\vec{E}_{ph}$ such that $\vec{E}_{ph} \perp \vec{E}_{e-h} \equiv \vec{E}_{ph} \cdot \vec{E}_{el} =0 $ i.e. neglegible coupling. The novel electron-TO phonon coupling mechanism is possible in the presence of an accompanying lattice distortion described by a generalized collective variable $\xi(x,t)$ (with its direction indicated by an orange dashed line). Inset: A slice of Fermi pocket showing a dispersion $\epsilon(k)$ in the vicinity of the heavy $x$ axis. In red we show displacement of electronic density localized around the $x$ axis due to the energy-momentum conservation that prohibits electron-phonon scattering processes away from the heaviest electron direction.}\label{fig:model}
\end{figure}

The assertion about coupling between electrons and the TO phonons is experimentally confirmed in an incipient FE, PbTe, by observation\cite{Fahy-photoexcit-experim} that i) photoexcited electrons do change the TO phonons frequency and ii) in return these TO phonons do modify the electrons' spectral gap. However, within the Fermi liquid framework it is assumed that the TO phonons are only very weakly coupled with electrons. This is based on the fact that the vector of polarization of the Fermi-liquid and the displacement vector associated with the TO mode are perpendicular to each other\citep{Herring-deform}
%(or almost, in lower symmetry lattices\citep{Herring-deform})
, which diminishes their coupling, see Fig.\ref{fig:model}. Even the most recent \emph{ab-initio} studies dedicated to PbTe\cite{Song-PbTe-abinit, Savic-repeated} conventionally claimed that the TO phonons have the weakest coupling with electrons\footnote{based  on a displacement potential scenario which is \emph{implicitly} equivalent to an advanced version of the above argument}. This has led to the contradictory conclusion that electron coupling with the TO phonons will play a very minor role close to the FE-QCP. Clearly, the TO coupling must emerge at a collective level, which is not captured in a single particle picture, and there is an obvious demand for theory to uncover the microscopic mechanism. %An unbiased proof of such a mechanism calls for a model that captures an energy transfer due to the emergent entanglement between electronic liquid and ionic motion. This requires a description 
An emergent model of correlated electron-ion dynamics in a non-adiabatic regime is needed to make an unbiased proof of the energy transfer. This is the key result of this work. 

The mechanism of electron-phonon coupling, present close to the FE-QCP, can be illustrated by a phenomenological argument (see Fig.\ref{fig:model}). An underlying reason of the TO phonon softening are long-range interactions between lattice vibrations. Such interactions change the character of the ion potential from quadratic to quartic\cite{Zhang-anharm-PbTe} along the line of FE distortion. In second quantization language $\hat{x}^4$ translates into phonon-phonon interactions. The TO phonon is not an eigen-state of the lattice oscillations any more, in its motion it is accompanied by a cloud of other low energy phonons\citep{Sangiorgio-Xray-broad}.% (an analogue of a polaronic cloud).
Physically this cloud of co-excited phonons (i.e. their displacements $\Xi_q$) is re-summed into a collective variable, a vector field $\vec{\xi}(x,t)$ that describes the ordering of the system in the vicinity of the QCP\citep{Devonshire-orig}\citep{Roussev-theoQPE-FE}. In the FE-QCP case the ordering is equivalent to an electric polarization of the system. Then, as illustrated in Fig.\ref{fig:model}, the field $\vec{\xi}(x,t)$ accompanying a TO mode induces an electric displacement field with a component in the direction of the electron liquid polarization. This configuration of induced local electric fields has a non-negligible energy, hence we have identified the new source of electron-phonon coupling. The strength of the electron-TO phonon coupling is proportional to the symmetry breaking $\vec{\xi}(x,t)$ which we consider as a dynamical variable. The aim of our study is to show that such a non-adiabatic electron-phonon coupling is not only non-negligible, but can give a substantial net effect on the transport properties.    

%\subsubsection{model (2-3 paragraphs)} 

We consider narrow gap, weakly doped semiconductors with extremely small carrier densities %The second component are TO phonons (these are the most affected by anharmonicity) which correspond to lattice displacement $\Xi$. 
whose hamiltonian is:
\begin{multline}\label{eq:ham}
H=\sum_{k}V_F k c_k^\dag c_k + \sum_{q}\omega_{\rm TO}(q)b_q^\dag b_q +\\
\sum_{k,q} V_{\rm ee}(k,q) c_{k1}^\dag c_{k2} c_{k3}^\dag c_{k4} \delta(k_1+k_3-k_2-k_4)|_{k=k_1+k_3, q=k_1-k_2}\\
+\sum_{q,q'}U_{\rm ah}(q,q') b_{q1}^\dag b_{q2} b_{q3}^\dag b_{q4}\delta(q_1+q_3-q_2-q_4)|_{q=q_1+q_3, q'=q_1-q_2}
\end{multline}
where the terms describe electron dispersion ($c_k^\dag$ operators), phonon dispersion($b_q^\dag$ operators), electron-electron interactions $V_{\rm ee}(k,q)$ and phonon-phonon interactions $U_{\rm ah}(q,q')$. The linear dispersion of fermions with velocity $V_F$ is justified in a narrow energy window (of order of a few THz) close to $E_F$. We expect the Fermi surface (FS) to be highly anisotropic, since the FS's with a high complexity factor\citep{Snyder-cmplx-thermoel} is common in TE, with a large effective mass in at least one direction and we study the dynamics along the $\hat{x}$ direction with the lowest value of $V_F$. %The $\omega^{TO}(q)$ is a dispersion of the TO mode which is responsible for the FE instability. 
The most prominent feature of our system is its proximity to FE-QCP (although on the quantum paraelectric side) and the parameters of the model must reflect this. A sole assumption of a long range nature of $U_{\rm ah}$\citep{OlleHellman-PbTe-anharm} (needed for FE) suffice\footnote{Supp.Matt} to strongly renormalize downwards $\omega_{TO}(q\rightarrow 0)$ and induce a large self-energy experimentally detected as broadening -- the waterfall effect\citep{Delaire-neutron-broad}. For larger $q$ the TO phonons become strongly dispersive and contribute substantially to transport. This is anomalous for this phonon branch and puts it beyond the adiabatic regime. %On the other hand, close to $\Gamma$ point, we have heavy phonons known (experimentally) to be extremely broadened i.e. they have large self-energy due to $U_{ah}$. 
A microscopic theory of pre-formed FE order parameter is derived by re-summing locally the $q\rightarrow 0$ vibrations to obtain a vector field of displacement $\vec{\xi}(x,t)$ which is proportional to an emergent polarization\cite{Devonshire-orig}. %The electron-electron interactions $V_{e-e}(k,q)$ have RPA character (dilute gas limit) but also contain an additional ingredient due to strong polarizability of ionic environment. The anharmonic term $U_{ah}(q,q')$ couples only TO phonons and is highly anisotropic with particularly high value along FE distortion direction(s). It's principal part has long-range dipole-dipole character (possibly with Frochlich correction) which means strong preference for forward scattering of phonons.      
Upon averaging out the low energy $\Xi_{q\rightarrow 0}$, see \footnote{Supp.Matt} for details, the effective hamiltonian for $T>\omega_{TO}(\Gamma)(\equiv \bar{\omega}_0)$ regime reads:
\begin{multline}\label{eq:ham-eff}
H=\sum_{k} V_F k c_k^\dag c_k + \sum_{q} \tilde{\omega}_r(q)b_q^\dag b_q + \sum_{k,q,k'} V_{\rm ee}(k,q) c_{k'}^\dag c_{k'-q} c_k^\dag c_{k+q}\\
 +\sum_{q} \tilde{g}_2 [\hat{\Xi}_q\cdot \tilde{\xi}(t)](b_{q}+b_{-q}^{\dag})(K\xi_c(t)+ M \ddot{\xi}_c(t))\\
 +\sum_{k,q} F_{el-ph}[k,q;\xi_c(t)][c_k^\dag c_{k+q}(b_{q}+b_{-q}^{\dag})]
\end{multline}  
where we take linear dispersion for the remaining \emph{rapid} phonons $\tilde{\omega}_r(q)=c_{ph}q$ and the last two terms, with $g_2 \equiv U_{\rm ah}|_{q_1\rightarrow 0}$ and $F_{el-ph} \sim \xi_c$,  are emergent due to the spontaneous symmetry breaking described by $\vec{\xi}(x,t)$. Terms with such an operator content have been derived using a different formalism in Ref.\cite{Horsfield_2004}. Here we take $\vec{\xi}(x,t)=\xi_c(t)\varphi(x)\otimes\tilde{\xi}$ with a spatial profile $\varphi(x)$ \footnote{Separation of spatial profile relies on the fact that there is equal number of left/right moving waves that combine into a standing wave.} known from QCP decay. The time dependence of the amplitude $\xi_c(t)$ is determined from an anharmonic equation of motion:
\begin{equation}\label{eq:EOM-class}
   \ddot{\xi}_c(t) + \tilde{\omega_0} \xi_c(t) + g (\xi_c(t))^3 = 0
\end{equation}   
Here $\tilde{\omega}_0=\sqrt{K/M} = min(\tilde{\omega}_r(q))$ and $g=U_{\rm ah}|_{x\rightarrow 0}$. The pseudo-spin variable $\tilde{\xi}(t)$, which captures the two possible directions of FE distortion, obeys the Ising-Kondo hamiltonian:
\begin{equation}\label{eq:Kondoish-xi}
H_{\tilde{\xi}}=\sum_q^{qh} \tilde{g}_2 \tilde{\xi}_z (b_{q}+b_{-q}^{\dag})
\end{equation}
Both Eq's.\ref{eq:EOM-class}-\ref{eq:Kondoish-xi} define analytically solvable theories. In the following, the parameters that enter into Eq.\ref{eq:ham-eff}-\ref{eq:Kondoish-xi}, are taken from Ref.\citep{Zhang-anharm-PbTe} and \citep{OlleHellman-PbTe-anharm} to make a link with a real material, PbTe. %The physics that we aim to capture is as follows. Due to the anharmonicity the TO phonon is accompanied by a cloud of low energy collective excitations. This is a bubble of pre-formed FE order with a finite polarization that may interact with electrons, hence $F_{el-ph}\sim \xi_c$. This coupling depends both on the relative momentum as well as the total momentum of the interacting particles. 
%However, since entire process is profoundly anharmonic, both components can mutually screen each other so to find an effective coupling one needs to investigate the time evolution of such strongly coupled system. 
The mutual motion of electrons and ions induces energy transfer between the two sub-systems, a central quantity of this study. To capture it we need to go beyond the adiabatic approximation. We then take a non-adiabatic case $V_F = c_{ph}$, which supports a maximal drag through vertical electron scatterings, i.e. electrons exchange energy with phonons and still remain available for further scattering events along the "heaviest" axis of the Fermi pocket instead of spreading evenly over the entire Fermi surface. The ability to exchange energy with phonons many times maximizes the distortion of occupancies along $\hat{x}$ (see Fig.1 inset).

%\subsubsection{the method}

The effective model given by Eq.'s \ref{eq:ham-eff}-\ref{eq:EOM-class} reduces our problem to electrons and phonons interacting with a classical variable whose equation of motion is known. %For electron-electron interaction we use random phase approximation (RPA), which is best in the dilute electron gas limit. To accommodate the influence of a dipole ($\xi_c$) field, in which electrons are moving, we generalize RPA and add the susceptibility of a dipole gas to the Lindhard susceptibility and include Singvi-Slojander local field corrections (i.e. vertex correction that should be included close to QCP).   
To solve the electron-phonon problem we use a variant of correlated electron-ion dynamics, where the analytic equations of motions for electron and phonon densities are integrated. The set includes the momentums' correlation, $\hat{\lambda}(q,t)$ in \footnote{Supp.Matt}, which is forced to be decoupled in the adiabatic approximation\citep{Horsfield_2004}, while in our time evolution it follows an unbiased trajectory caused by $F_{el-ph}$. The simpler, real-space variant of the method, called ECEID\citep{Val-thes}, was proven to be able to capture non-adiabatic effects beyond Ehrenfest dynamics\cite{ECEID}. In the EOM for the new $\hat{\lambda}(q,t)$ the two body density matrix for electrons enters. It is here where the $V_{\rm ee}$ term intervenes and we tackle it in the random phase approximation (RPA) generalized to add the susceptibility of a dipole gas to the Lindhard susceptibility and to include the Singwi-Slojander local field corrections.  %It should be pointed that originally the method was developed in a real space setting, but is our case -- dedicated to weakly doped semiconductors -- the reciprocal space has to be used. 
%Once the time evolution is established one can use fluctuation-dissipation theorem: the Fourier transform of the relaxation gives an access to frequency dependent density-density susceptibilities of the system. 
%\subsubsection{results1: electron-phonon scattering rate}
%We first investigate an event of creation of a single TO phonon and its impact on the electron system.  In this way, by studying the response of electronic density we shall obtain the effective electron-phonon coupling matrix. 
%The ECEID method provides us with a time evolution of the densities for coupled electron-phonon system starting with a given initial condition. 
We develop our results in two steps: first we compute the effective electron-phonon coupling and then we use it to estimate the transport coefficient.
 
\emph{Electron-TO phonon coupling} The initial state is the thermal equilibrium state $|\Psi\rangle_T$ with one  phonon added (in a given q-mode)\footnote{we are concerned with finite temperature properties, hence an operator $b_q^\dag$ is applied not to a ground state, but to an equilibrium, thermal state at given finite temperature $T$}: $|\Psi\rangle_i = b_q^\dag(t=t_0)|\Psi\rangle_T$. We study the expectation values of equal-time fermionic occupancies $\langle c^{\dag}(t,k)c(t,k)\rangle$ and by monitoring their time dependence, as the extra phonon is propagating through the system, we are able to obtain the transferred energy $\delta E_k[|\Psi\rangle_i] \sim \delta n_k [|\Psi\rangle_i]$ which is equal to the electron-phonon coupling strength. %To be precise, the bare dipole-dipole interaction is an input parameter of the hamiltonian, while as a result of calculation we get a quantity renormalized by electron-electron and electron-phonon interactions.

\begin{figure}[t]
\centering
 \includegraphics[width=0.9\linewidth]{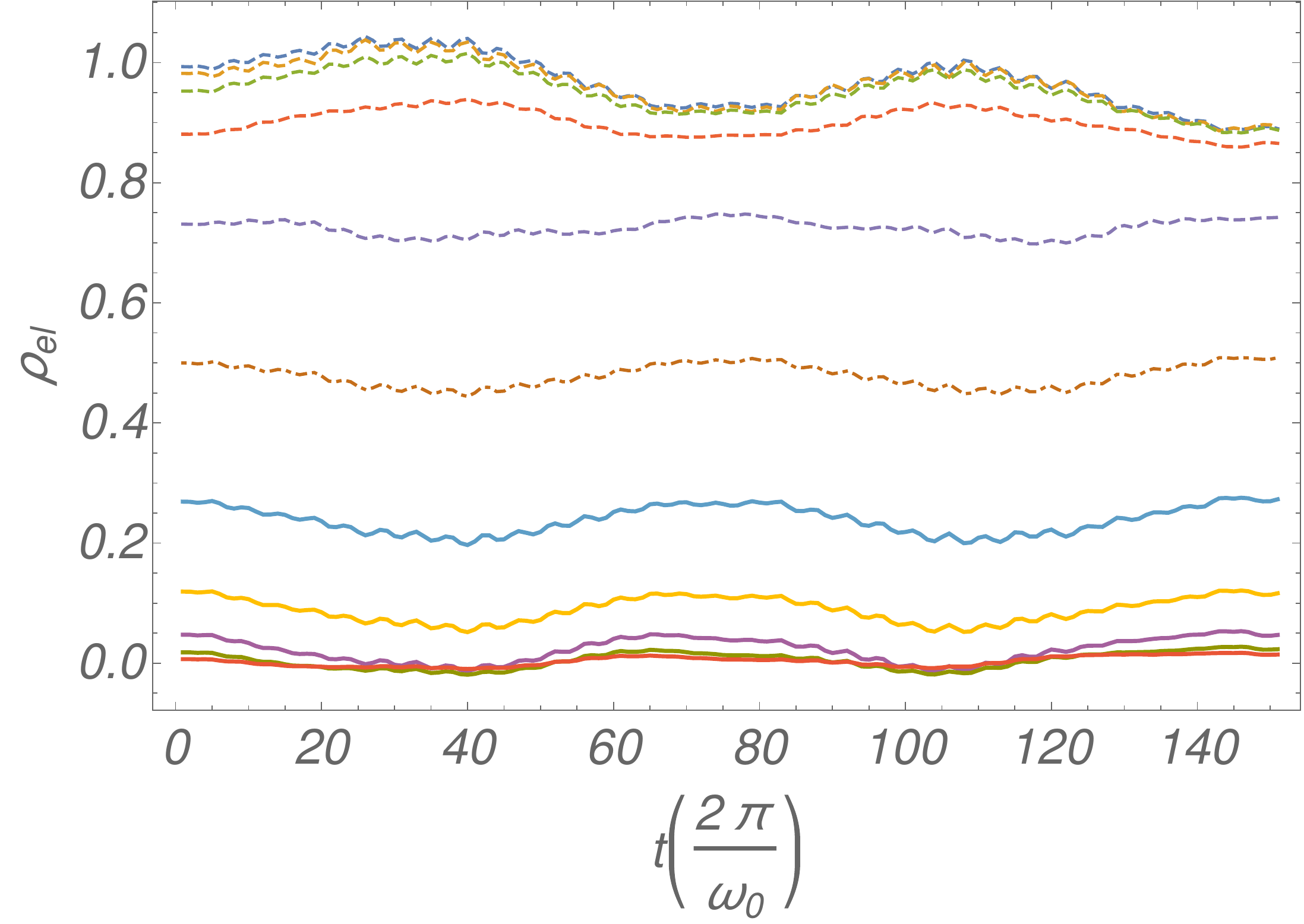}\\
 \includegraphics[width=0.9\linewidth]{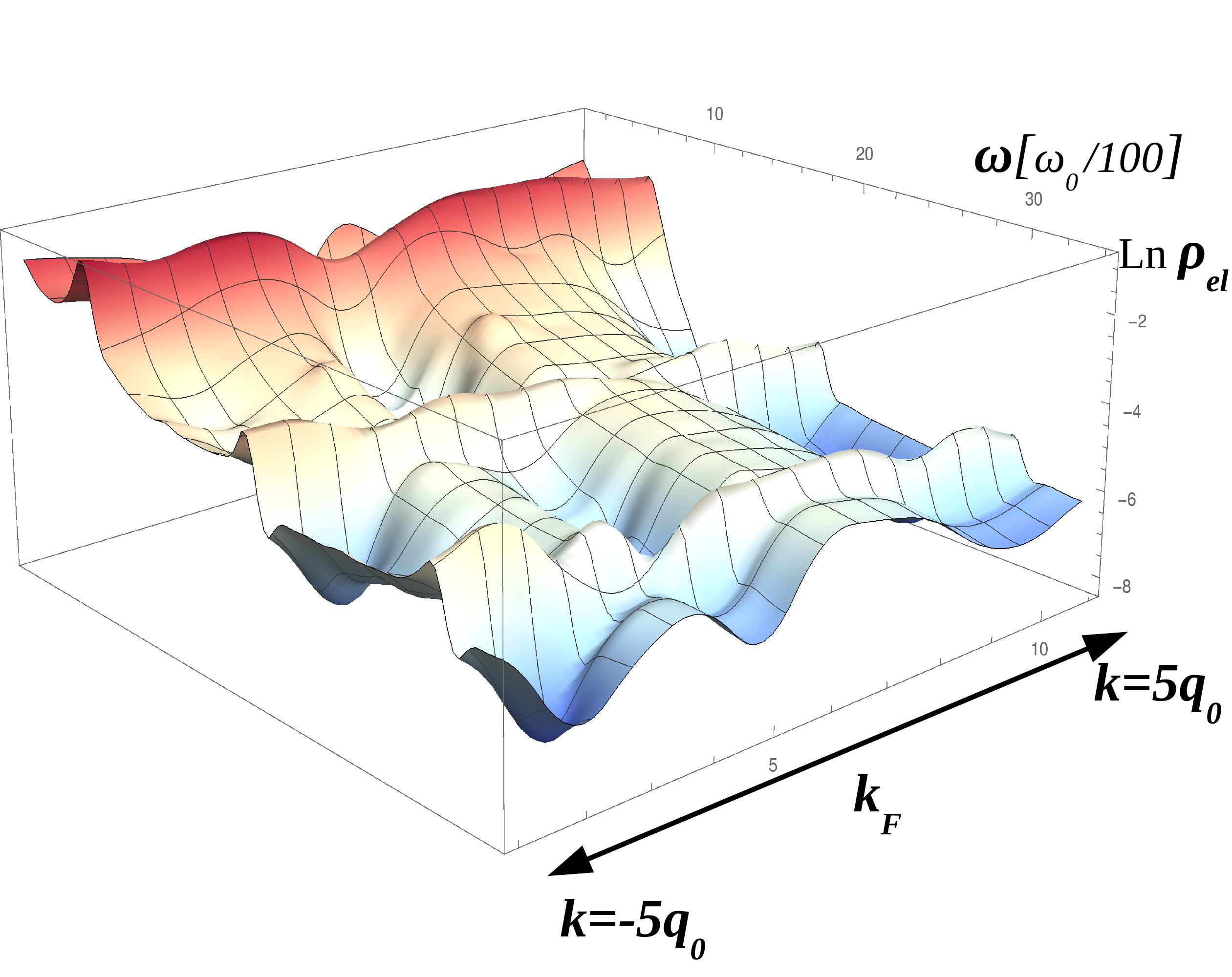}
  \caption{Response of the fermionic system upon applying a bosonic creation operator in a $q_1=1 q_0$ mode with energy $1\tilde{\omega}_0$. The calculation is done for discrete bosonic/fermionic modes with energies $E_n=n\omega_0$ (for fermions $\epsilon=E_F\pm E_n$) where $\omega_0$ is the lowest energy of the linear part of TO spectrum (hence around 20\% above $\bar{\omega_0}$), $n \in [1, 5]$.  In the top panel the temporal evolution of expectation values of fermions occupancies are shown for each bosonic mode, this is a solution of our EOM. In the bottom panel we show its Fourier transform (on a natural logarithmic scale), hence an effective electron-TO phonon coupling as a function of energy and (fermion's) momentum.}
  \label{fig:response-n}
\end{figure}

The results of the calculation are shown in Fig.\ref{fig:response-n}. Our simulations clearly show the presence of electron-phonon coupling -- there is an energy transfer due to an extra phonon into the fermionic system. 
%Initially, the phonon occupancies vary strongly, as an exponential function of time, as the . As time progresses this initial relaxation is superposed with a quasi-stationary state and.
The electronic densities $\hat{\rho_e}$ are not constant, their time evolution is defined by high and low frequency oscillations. We are modelling a closed quantum system, so we monitor how the energy is transferred between the available discrete levels. In order to analyse the oscillations we take a Fourier transform of the time dependence. Since the transient state, when phonons are trying to accommodate the extra excitation, is observed only for phonon densities, the spectrum of quasi-stationary $\hat{\rho_e}$ reveals the clear separation of energy scales. The high frequencies can now be associated with the difference of discrete energy levels of the harmonic model (with quantized values of momenta) and are broadened by anharmonicity. At low frequencies we observe the most pronounced peak whose energy scale falls close to $F_{el-ph}$. We interpret this feature as being present due to Rabi oscillations that develop in our closed quantum system that was initially set out of equilibrium. In this interpretation the excited phonon sub-system exerts its influence on discrete occupations of fermionic levels while the quasi-classical coupling is sufficient to drive the Rabi oscillations. What we observe is in fact a dynamics of respective Floquet states. %Then the position of the peak (upon initially exciting various phonon q-modes) informs us about an effective q-dependence of the coupling, while
The varying amplitude of oscillation for various fermionic states informs us about the scattering dependence on fermion momentum.% at a distance from Fermi level.

%results 1 (1 page): we perform time evolutions and then (by Fourier transform) inspect energy/momentum resolved (for each single athermally occupied phonon mode we have an energy spectrum) lesser Green's function of phonons (to show mode-mode coupling) and lesser Green function of electrons (to show the presence of a dynamic electron-phonon coupling).

%\begin{widetext}
%\onecolumngrid
%\twocolumngrid
%\end{widetext}

To explore the quantitative effect of the FE-QCP proximity, we need to find how the parameters in Eq.\ref{eq:EOM-class} are changing as the system is moved on the $T-\delta$ plane (here $\delta$ is a control parameter e.g. strain). The renormalization group flow of scalar $\phi^4$ theory\citep{Hertz76-QCP} is known to give a correct description of the FE-QCP\citep{Roussev-theoQPE-FE}\citep{Rowley-FEQCP}, then we know universal flow of the Gruneisen parameter\citep{Gruneis-scaling} and its relation with the anharmonicity\cite{anharm-to-Gruneis}. 

%\subsubsection{results2: thermoelectric transport coefficient}

\emph{Transport coefficients} We focus on the Seebeck coefficient which can be expressed in terms of Kubo current-current susceptibilities $L^{ij}$: $S=L^{\sigma,\kappa}/(T L^{\sigma,\sigma})$ where $L^{\sigma,\kappa}$ will contain both electronic liquid dissipative part $L^{\sigma,\kappa}_{el}$ and new phonon-drag part $L^{\sigma,\kappa}_d$. As the system approaches the FE-QCP it is known that the electronic resistivity ($\sim 1/L^{\sigma,\sigma}$) increases but at the same time the Fermi pockets shrink so the entropy associated with these carriers ($\sim L^{\sigma,\kappa}_{el}$) also decreases. We then aim to compute the drag component $L^{\sigma,\kappa}_d$ to check if it is able to profit from QCP fluctuations of $\xi_c$. 

 %In a Kubo formalism (linear response), in a local limit, this gives desired off-diagonal transport coefficient:   

%\begin{equation}
%L_{\sigma,\kappa}(t,t')=\langle [j_\sigma(t), j_\kappa(t-t')] \rangle
%\end{equation}

%The Fourier transform of the correlator of expectation values gives us $L^{\sigma,\kappa}_d(\omega)$. 
From the time evolution of the densities $\Delta\rho_{e}(\epsilon_i,t)$ one can extract expectation values of electron and phonon currents.% from their respective continuity equations (Gauss theorem).This allows us to 
and compute cross-correlation functions of currents' variations\footnote{strictly speaking one could also recursively repeat procedure to compute 2nd, 3rd, etc  order approximations for the displaced phonon distribution. Here we assume that linear response holds provided our initial 1st order choice $\Delta N_b (T)$ is close enough to the real displacement.}.
Extra care needs to be taken, since the output of our simulation is a superposition of Floquet states. %(note that at finite distance from QCP there exists a characteristic frequency associated with the rattling motion). 
However, one may argue that the drag effect itself takes place in the presence of unequilibrated %(although stationary)
state of the phonon distribution $N_B^{(0)}+\Delta N_B$. Hence our description captures the experimental situation rather well. We then apply Kubo formalism adapted, by Lehmann representation of currents cross-correlation functions $\langle j_\kappa^j \rangle$, for the Floquet states\cite{Oka-Floq-cond1}
\begin{equation}\label{eq:Floq-conductivity}
L^{\sigma,\kappa}= \imath \sum_{i,j,\omega_n<T}\frac{n_F(i)-n_F(j)}{\epsilon_i-\epsilon_j}\frac{\langle\langle j_\sigma^i \rangle\rangle\langle\langle j_\kappa^j \rangle\rangle}{\epsilon_i-\epsilon_j+\omega_{n}+\imath\delta^{+}}
\end{equation}
where $n_F(i)$ is a Fermi-Dirac distribution for a state with energy $\epsilon_i$, the summation is over the Floquet states (oscillation index $n$), over all available modes $i,j$  and the double average indicates average in time and over all possible initial/final states in an energy window around given discrete level.% (constant density of states for a linear dispersion). 

\begin{figure}
 \includegraphics[width=0.9\columnwidth]{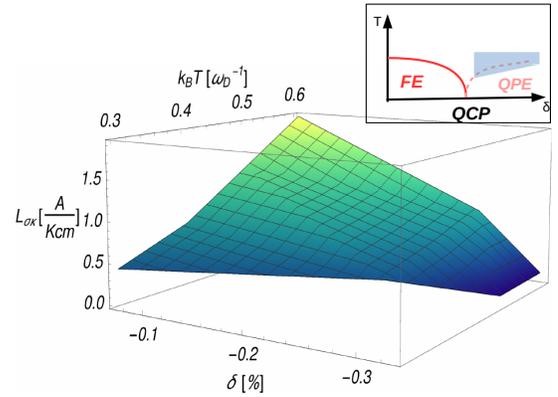}
  \caption{Thermoelectric conductance $L^{\sigma,\kappa}_d$ as the system approaches the FE-QCP, where the QCPs vicinity is parameterized as a function of strain and temperature. The natural unit of thermoelectric conductance is $\hbar e / k_B $, natural length-scale is $2\pi/q_0$ and energy unit throughout the calculation is $\omega_0$. By taking $\omega_0^{PbTe}|_{\delta=-0.3\%}=1THz$ it allows us to estimate the amplitude of the effect on a vertical scale. The inset shows the region in the parameter space (grey) where our reasoning applies. PbTe would be located on the right side of the grey zone.}
  \label{fig:Seebeck-coeff}
\end{figure}

The result of this procedure is shown in Fig.\ref{fig:Seebeck-coeff}. The temporal fluctuations of both densities can produce a net correlation between respective currents, hence the drag effect.  %Furthermore, on the bottom panel, we see that 
The effect becomes more pronounced as the system approaches the QCP, where the relative effect of $\xi_c$ fluctuations is enhanced. This should be compared with a maximal $L^{\sigma,\kappa}$ measured in PbTe, $L^{\sigma,\kappa}$=0.5 A/(Kcm), which indicates that the novel effect proposed here may contribute significantly. Our plot is restricted to a regime where our approximation applies (inset). As we increase temperature, we shall ultimately enter into a regime, not captured in our \emph{minimal} model, where the thermal fluctuations destroy the temporal coherence of $\xi_c$ \citep{Coleman-FE-Casimir}%(to be precise the power law decay function turns into Bessel function of the second kind) 
and furthermore, in case of PbTe, relaxation into LA modes will appear\citep{Delaire-neutron-broad}. Hence the increase of $L^{\sigma, \kappa}$ should be suppressed when $T \gtrapprox \omega_D$ \footnote{$\omega_D$ is the Debye frequency which in the case of PbTe it is $6$ to $8$ times larger than $\omega_0$}. For all points on the plot in Fig.\ref{fig:Seebeck-coeff} we have used the same initial condition $\Delta N_{B}(q)$ proportional to polylogarithmic functions\footnote{Supp.Mat.} so the observed $T-\delta$ dependence comes from the critical dynamics of the system, not from a biased choice of the initial condition. The changes of $L^{\sigma, \kappa}$ in Fig.\ref{fig:Seebeck-coeff} are solely due to modified parameters of the EOM, which in turn follow modifications (of e.g. $g_2$\citep{Gruneis-scaling}) known from the $\phi^4-$QCP theory\citep{Hertz76-QCP}. The increase of thermoelectric conductivity (drag) is exclusively due to increased anharmonicity, as duly captured in our model. %\emph{However the occupancy of phonons may change also as a function of $\delta$, steming from an energy dependence of self-energy. By means of Ward identity it can be shown that such dependence must be accompanied by incorporating vertex correction to electron-phonon coupling, it is a known breakdown of single-particle perturbative picture close to QCP.  These effects can be accounted for by an extension of our formalism -- we have computed the equal-time lesser Greens functions which can be transformed into retarded Greens function and then upon integrating out momenta/energy produce self-energy for a given mode -- this calculation will be posponed for a subsequent publication.}  

%results 2 (1 page): cross-correlation of electronic and phononic currents leads to $L^{12}(\omega)$ and $L^{11}(\omega)$ Kubo coefficients. Then the Seebeck coefficient is [Mahan]: $S=L^{12}/(T L^{11}) + d Log[L^{11}(\omega)]/d \omega$  where the second term is a dissipative component estimated through the Mott relation for Fermi liquid.[Note: when computing the Kubo terms we will need to introduce an artificial imaginary self-energies $\equiv$ relaxation times through 'other' channels (electron-LA phonon, electron-impurity, electron-electron). Otherwise we would have infinitely long lasting currents.]

%An issue A: strictly speaking the $L_{12}$ depends on preparation of initial state i.e. one has to take most likely distribution of bosons out of equilibrium. To this end I would propose to use a solution of a linearized Boltzman equation with relaxation time at each q known because we know el-ph coupling vertexes. Questions: i) how to estimate the total relaxation time (that I will also use to get the final response from cross-correlator)? ii) what is the role of phonons' mode-mode mixing?

%An issue B: the $L_{11}$ enters expressions above and here one has to estimate it somehow. There will be two contributions: i) from el-el interactions (note that I have an approximation for charge susceptibility) and ii) from el-ph interaction (proportional to the vertex squared).  

%Before we conclude, we wish to emphasize the broader importance of our theoretical findings. For nearly a century the adiabatic assumption, where electrons immediately adjust to the motion of ions, has been taken as a standard.
The broad importance of this work stems from our method that goes beyond adiabatic approximation -- a paradigm, since Born-Oppenheimer work\cite{BO1927} almost a century ago, that has turned out to be insufficient with the advent of theoretical\citep{Oka-Floq-cond1, RevModPhysGiustino} and experimental\cite{Novosiel-graph-nonadia} quantum non-adiabatic and non-equilibrium dynamics. %In a broad field of molecular and nano-physics this is apparent in the situation of while in the solid state the need to go beyond is apparent in the vicinity of any phase transitions where lattice displacement drastically changes dielectric constants. 
Formally, the adiabatic ansatz means that the Fock spaces for many body bosonic and fermionic states are decoupled into a simple product, which implies that their entropies are additive. In practice, applying the phonon creation operator (in the adiabatic basis) does not alter the many body state of fermions. Our formalism then clearly allows to go beyond this approximation: in Fig.2 we directly observe a change of electronic density of states  $\Delta \hat{\rho}_{\rm e}$ \emph{after} $b_q^\dag|\Psi_G\rangle$. The \emph{phonon-induced} off-diagonal terms $\langle \Psi_{\rm +q}|b_q^\dag|\Psi_G\rangle = t_{\rm +q-G}\neq 0$ are precisely those neglected in the adiabatic approximation. Our result provides a direct access to a measure of non-adiabaticity in the system and furthermore, through $\hat{\rho}_e=\hat{\rho}_e^{(0)}+\Delta \hat{\rho}_{\rm e}$, all electronic observables can be computed. Here we have chosen to focus on one quantity, the thermoelectric conductivity, which is directly proportional to the entropy transfer between the two sub-systems. Furthermore, we choose the simplest model, with a linear coupling to the TO mode (otherwise prohibited in the high symmetry structure) and all other ingredients (critical dynamics, anharmonic motion) obey exact analytical solutions. This is for the sake of transparency, to study a situation where non-adiabatic phenomena manifest particularly clearly. %Yet the situation pertains to describe the the very relevant situation of  distortive FE phase transition,  such that the only unknown comes from non-adiabaticity.   

In conclusion we have shown that close to a FE-QCP strong anharmonic fluctuations of atomic positions, described by a collective variable $\xi_c$, lead to a new type of coupling between electrons and anharmonic phonons. It is a strong coupling effect that one is able to detect only by going not only beyond the standard adiabatic approximation for electrons but also beyond Ehrenfest dynamics for ionic motion. Our model is applicable in any distortive QCP where electronic polarizabilities are modified. Furthermore we have shown that the coupling through $\xi_c$ can produce a finite net phonon drag effect, which substantially enhances thermoelectric transport coefficients. %Such situation is particularly relevant in weakly diluted narrow gap semiconductors, where close to band bottom the Fermi velocity may be very small.
A superconducting phase, in an analogue material STO, is present even closer to the QCP and therein the source of a strong electron-TO phonon coupling remains an mystery. A possible extension of this work is to cover this intriguing regime. % and we hope that an extension of our mechanism could provide an explanation. 

\emph{Acknowledgements} I am greatly indebted to Sucismita C. Chutia and Jorge Kohanoff for a careful reading of the manuscript and many valuable remarks. I also wish to thank T.Todorov and P.Aguado Puente for many helpful discussions. This work was supported by a research grant from Department for the Economy Northern Ireland and Science Foundation Ireland (SFI) under the SFI-DfE Investigators Programme Partnership, Grant Number 15/IA/3160.

%conclusion2: Entire reasoning here is based on an assumption that long-range anharmonicy is present. This picture is fully self-consistent, but the issue is what triggers it? It is known that softening of TO also closes the band gap [Fahy, Nat.Comm.] so it means more carriers that screens any incipient dipole. The only way to avoid it is by assuming some finite coupling between TO phonons and electrons that can re-shuffle dilute electron liquid. Recent paper identified regimes where such coupling may be present and they are actually met in PbTe: spin-orbit coupling (Pb hosts unpaired electronic charge which varies with distortion) and changes of orbital character (when TO mode aquires dispersion then it links electrons at various energies away from $E_F$ and the peculiarity of PbTe is that orbital content changes as we vary eigen-energy). To test this we have added a small constant el-ph coupling on the top of our displacement driven coupling and found that the displacement of electronic density is stabilized and enlarged. Hence the two mechanisms cooperate to push the system closer to FE.   

%conclusion3: although our interest here was in a transport phenomena that takes place at energies at $\approx 100K$ our results shall be also applicable in a regime located much closer to QCP. With this respect we point out to a recent interest of a broad community in very similar problem of el-TO phonon coupling in very weakly doped STO, where strong coupling superconductivity has been detected. We hope our work will inspire future studies in this field as well.

\bibliography{PbTe-biblio}  
 
\end{document}